\documentclass[a4paper,11pt]{article}

\usepackage{jheppub} 
\usepackage[T1]{fontenc}
\usepackage{graphicx}

\newcommand{\bt}{\mbox{\boldmath $\theta$}}

\newcommand{\bx}{\mbox{\boldmath $x$}}

\title{\boldmath Hessian potential for Fefferman-Graham metric}

\author{Hiroaki Matsueda}
 
\affiliation{Sendai National College of Technology, Sendai 989-3128, Japan}

\emailAdd{matsueda@sendai-nct.ac.jp}

\abstract{
The Fefferman-Graham metric is frequently used for derivation of the first law of the entanglement thermodynamics. On ther other hand, the entanglement thermodynamics is well formulated by the Hessian geometry. The aim of this work is to relate them with each other by finding the corresponding Hessian potential. We find that the deformation of the bulk Hessian potential for the pure AdS spacetime behaves as a source potential of the boundary Fisher metric, and the deformation coincides with the Fefferman-Graham metric. A peculiar feature different from related works is that we need not to use the Ryu-Takayanagi formula for the above derivation. The canonical parameter space in the Hessian geometry is a kind of the model parameter space, rather than the real classical spacetime in the usual setup of the AdS/CFT correspondence. However, the underlying mathematical structure is the same as that of the AdS/CFT correspondence. This suggests the presence of more global class of holographic transformation.
}


\begin{document}

\maketitle

\flushbottom

\section{Introduction}

In a series of my recent papers~\cite{Matsueda1,Matsueda2,Matsueda3}, we have been developing the entanglement thermodynamics in terms of the Hessian geometry~\cite{Casini,Guo,Blanco,Alishahiha,Wong,Takayanagi2,Takayanagi3,Takayanagi4,Faulkner,Nima,Banerjee,Arpan1,Arpan2}. The entanglement thermodynamics is one of important topics in the AdS/CFT correspondence~\cite{Maldacena1,Maldacena2}, and thus much deeper interpretation from various viewpoints is required. Since the information-geometrical analysis of entropy has been also interested in very wide interdisciplinary research fields~\cite{Amari,Shima,Balian,Barbaresco}, the present work would facilitate the continuous development of these fields. My approach is rather different from the standard ones in which one usually relys on the Ryu-Takayanagi formula for the calculation of the holographic entanglement entropy~\cite{Takayanagi1}. Therefore, we would obtain alternative viewpoints for both of the entanglement thermodynamics and the Ryu-Takayanagi formula. In the previous approaches~\cite{Casini,Guo,Blanco,Alishahiha,Wong,Takayanagi2,Takayanagi3,Takayanagi4,Faulkner,Nima,Banerjee,Arpan1,Arpan2}, an underlying motivation would originate in interdesciplinary relationship among thermodynamic features of the Einstein equation and information-theoretical aspects of black holes. On the other hand, my approach can be interpreted as reduction from (microscopic) statistical mechanics for the entanglement Hamiltonian. How these two merge together is an interesting question. The important point here is that the structure of the Hessian potential contains rich information of the area law scaling in a holographic calculation of the entanglement entropy. Since in the previous papers we have already obtained the general formula of the entanglement thermodynamics in the Hessian-potenial approach, the purpose of this study is to transform it into a new form that is closely related to the standard representation.

Up to now, in the quantum field theory side, the following results have been obtained. Let us start with the asymptotically AdS${}_{d+2}$ background
\begin{eqnarray}
ds^{2}=\frac{R^{2}}{z^{2}}\left[-f(z)dt^{2}+\frac{1}{f(z)}dz^{2}+\sum_{i=1}^{d}(dx^{i})^{2}\right],
\end{eqnarray}
where $\bx=(x^{0},x^{1},...,x^{d})$, $x^{0}=t$, $f(z)\simeq 1-mz^{d+1}$, and $R$ denotes the curvature radius of AdS. A perturbation from pure AdS${}_{d+2}$ metric is also treated by the Fefferman-Graham (FG) gauge
\begin{eqnarray}
ds^{2}=R^{2}\frac{dz^{2}+g_{ij}(z,\bx)dx^{i}dx^{j}}{z^{2}}.
\end{eqnarray}
Here, the metric perturbation is represented as
\begin{eqnarray}
g_{\mu\nu}=\eta_{\mu\nu}+h_{\mu\nu}.
\end{eqnarray}
Thus $h_{\mu\nu}$ induces the perturbation to the flat Minkowski spacetime $M^{1,d}$ where the CFT${}_{d+1}$ lives. We assume that $h_{\mu\nu}$ is small and take the first order perturbation. The dynamics by this perturbation is described by the Einstein equation with a negative cosmological constant. This means that the excitations and the energy-momentum tensor in the CFT side are characterized by $h_{\mu\nu}$.

By the above setup, we can derive the following equality similar to the first law of thermodynamics~\cite{Casini,Guo,Blanco,Alishahiha,Wong,Takayanagi2,Takayanagi3,Takayanagi4,Faulkner,Nima,Banerjee,Arpan1,Arpan2}
\begin{eqnarray}
T_{\rm eff}\Delta S_{A}=\Delta E_{A},
\end{eqnarray}
where we consider a special region $A$, called entangling surface, at the AdS boundary, $\Delta S_{A}$ is the difference of the entanglment entropy between the ground and excited states, $\Delta E_{A}$ denotes the corresponding energy difference, and $T_{\rm eff}$ is the entanglement temperature. Both of $\Delta S_{A}$ and $\Delta E_{A}$ are described in the holographic terminology, and thus we can directly compare them with each other by using some geometric objects. At first, the variation of the entropy, $\Delta S_{A}$, is derived from the Ryu-Takayanagi holographic formula, $\Delta S_{A}=\Delta\gamma_{A}/4G$, by evaluating the change in the minimal surface $\Delta\gamma_{A}$. When the region $A$ is ball-shaped with radius $l$, the surface $\gamma_{A}$ specified with $r=r(z)$ is defined by
\begin{eqnarray}
\gamma_{A}=R^{d}\Omega_{d-1}\int_{\epsilon}^{u}\frac{dz}{z^{d}}r(z)^{d-1}\sqrt{g(z)+r^{\prime}(z)^{2}}.
\end{eqnarray}
From this definition, we can take the variation under some assumptions. The entanglement temperature takes the universal value $T_{\rm eff}=(d+2)/2\pi l$. On the other hand, the total energy $\Delta E_{A}$ in a region $A$ is evaluated as $\Delta E_{A}=\int_{A}d^{d}x T_{00}$ in which $T_{00}$ is the energy density in CFT${}_{d+1}$. When we expand $h_{\mu\nu}$ as
\begin{eqnarray}
h_{\mu\nu}=z^{d+1}H_{\mu\nu}+\cdots,
\end{eqnarray}
in the near AdS boundary $z\rightarrow 0$, the holographic stress tensor is obtained as~\cite{Balasubramanian,Haro}
\begin{eqnarray}
T_{\mu\nu}=\frac{(d+1)R^{d}}{16\pi G_{N}}H_{\mu\nu}. \label{TH}
\end{eqnarray}
By using this relation, we can represent $\Delta E_{A}$ holographically. Thus, both of $\Delta S_{A}$ and $\Delta E_{A}$ are represented by the metric perturbation.

A purpose of this work is to reconstruct the above results in terms the Hessian geometry, without the Ryu-Takayanagi formula. If this statement is correct, this means that the above peculiar features have been already contained in the theory of the Hessian geometry. For the precise determination of the metric perturbation in the Hessian geometry, it is crucial to find the Hessian potential that exactly leads to the FG metric. We will actually find such potential form and will derive that the deformation of the bulk Hessian potential for the pure AdS spacetime behaves as a source potential of the boundary Fisher metric. At the same time as already mentioned, our basic setup in both of the AdS and CFT sides looks quite different from the standard approaches. Thus, we emphasize mathemetical similarity and some essential differences between them. This indicates that there might exist more global structure that unifies various types of the AdS/CFT-type correspondences~\cite{Blau,Shock}.

The organization of this paper is as follows. In Sec.~II, we briefly summarize the Hessian-potential approach for the AdS/CFT correspondence. In Sec.~III, we introduce a Hessian potential containing a new function that is finally converted to the FG term. We examine the properties of this term in both of classical and quantum sides, and show that the present formulation is mathematically quite similar to the previous results. In the final section, we will summarize the results obtained in this paper.

\section{Hessian-potential formulation of entanglement thermodynamics}

\subsection{Information-geometrical interpretation of AdS/CFT}

Let us explain the method of the Hessian-potential approach to the information-geometrical interpretation of the AdS${}_{d+2}$/CFT${}_{d+1}$ correspondence. Suppose there exists a Hessian potential $\psi$ which is a function of $(d+2)$-canonical parameters $\bt=(\theta^{0},\theta^{1},...,\theta^{d+1})$. This potential is a source of producing both of the Fisher metric in the $(d+2)$-dimensional classical spacetime by $\bt$ and the entanglement entropy in a quantum field theory. 
In other words, one of advantageous points of the present approach is that we can simply unify the both sides by the same framework. In the quantum side, the $(d+2)$-canonical parameters are the model parameters, and then $\theta^{d+1}=\theta$ is attributed to a scale parameter necessarily introduced by the truncation of environmental degrees of freedom in the theory of entanglement as we will soon explain it. The $(d+1)$-parameters, the rest of $\theta$, are for example filling fraction, time, and so on. In general, it is not deterministic whether the spacetime dimension of the quantum side is $d+1$ or not. Furthermore, the canonical parameter space is a classical one, but the coordinates may not originate in the real spacetime coordinates in the quantum side. Thus, the basic setup seems to be different from usual AdS/CFT. Thus, we might obtain a viewpoint that enables us to access more global class of holography. Fortunately, in the lattice free fermion model with $d=1$, the number of the canonical parameters are two except for $\theta$, and one of $\bt$ is time, and then the setup is similar to the usual AdS/CFT.

At first, the Fisher metric is defined by
\begin{eqnarray}
g_{\mu\nu}(\bt)=\partial_{\mu}\partial_{\nu}\psi(\bt),
\end{eqnarray}
where $\partial_{\mu}=\partial/\partial\theta^{\mu}$ and the Greek index $\mu$ runs from $0$ to $d+1$. We will abbreviate $\partial_{d+1}$ as $\partial$. Let us next introduce a quantum state in the Schmidt decomposition form
\begin{eqnarray}
\left|\psi\right>=\sum_{n}\sqrt{\lambda_{n}(\bt)}\left|A;n\right>\otimes\left|\bar{A};n\right>, \label{Schmidt}
\end{eqnarray}
where the Schmidt coefficients $\lambda_{n}$ depend on the canonical parameters, since $\bt$ are model parameters. Now, we think the entanglement between two subsystems $A$ with linear size $L$ and $\bar{A}$, and the trucation of the environmental degrees of freedom induces mixed-state feature or thermal properties. Thus, it is quite natural to think that the coefficients are in the exponential family form as
\begin{eqnarray}
\lambda_{n}(\bt)=\exp\left(\theta^{\alpha}F_{n\alpha}-\psi(\bt)\right)=\frac{1}{Z}\exp\left\{-(-\theta^{\alpha}F_{n\alpha})\right\}=\frac{1}{Z}e^{-E_{n}/T_{\rm eff}}, \label{thermal}
\end{eqnarray}
where $\psi=\log Z$, $T_{\rm eff}$ is the entanglement temperature and $E_{n}/T_{\rm eff}=-\theta^{\alpha}F_{n\alpha}$. It is necessary to examine whether this conjecture is justified. At least for a free fermion model, this conjecture is correct~\cite{Matsueda1,Matsueda2}.

According to the entanglement thermodynamics in the Hessian-potential representation, the entanglement entropy of the quantum system is represented as
\begin{eqnarray}
S(\bt)=\psi(\bt)-\theta^{\alpha}\sum_{n}\lambda_{n}F_{n\alpha}=\psi(\bt)-\theta^{\alpha}\partial_{\alpha}\psi(\bt). \label{exact}
\end{eqnarray}
When we define the entanglement free energy as $F=-T_{\rm eff}\ln Z$ and the entanglement energy as $E=\sum_{n}\lambda_{n}E_{n}$, we find
\begin{eqnarray}
F=E-T_{\rm eff}S.
\end{eqnarray}
Therefore, the macroscopic property derived from averaging over the microscopic degrees of freedom is equivalent to the law of standard thermodynamics. In that sense, it would be better to say that my approach is the entanglement statistical mechanics. Furthermore, if we take the derivative by $\theta^{\beta}$, we find
\begin{eqnarray}
\partial_{\beta}S(\bt)=-\theta^{\alpha}\partial_{\alpha}\partial_{\beta}\psi(\bt)=-\theta^{\alpha}\partial_{\beta}\eta_{\alpha},
\end{eqnarray}
where $\eta_{\alpha}=\partial_{\alpha}\psi$ is a Legendre parameter. This leads to the differential form of the first law of thermodynamics
\begin{eqnarray}
dS=-\theta^{\alpha}d\eta_{\alpha}. \label{differential}
\end{eqnarray}
This is also a thermodynamic formula basically equivalent to $T_{\rm eff}\Delta S_{A}=\Delta E_{A}$, since the expectation value of the entanglement Hamiltonian is given by $E=T_{\rm eff}(-\theta^{\alpha}\eta_{\alpha})$. Here, the variation is taken for the Legendre parameters. According to the linear response theory, $\eta_{\alpha}$ is a fource field by the Hessian potential, and thus $\theta^{\alpha}$ is induced flow. In the present case, Eq.~(\ref{differential}) tells us that $\theta^{\alpha}$ is information flow by $\psi$.

The violation of the volume-law scaling of the entanglement entropy is characterized by the information flow in the holographic side. In we can define an entangling surface by an appropriate way, we can ask what kind of information flow occurs across the surface and how this is related to the holographic representation of the entanglement entropy. For the above purpose, it is convenient to calculate the covariant derivative of the canonical parameters as
\begin{eqnarray}
\nabla_{\alpha}\theta^{\alpha}=\partial_{\alpha}\theta^{\alpha}+\Gamma^{\alpha}_{\;\alpha\beta}\theta^{\beta}=-\frac{1}{2}g^{\alpha\beta}\partial_{\alpha}\partial_{\beta}\left(S-\psi\right),
\end{eqnarray}
where the Christoffel symbol is described as $\Gamma^{\lambda}_{\;\mu\nu}=(1/2)g^{\lambda\tau}\partial_{\tau}\partial_{\mu}\partial_{\nu}\psi(\bx)$. As we will later see the case of the AdS/CFT correspondence, we can indentify $S$ with $\psi$ after the second derivative of them by the canonical parameters. Then, this quantity almost vanishes. To look at this feature as well as an appropriate definition of the entangling surface, we first assume the presence of an entangling surface $\Sigma$ which has rotational symmetry along an axis and $\eta_{\alpha}$ is proportional to a unit vector $n_{\alpha}$ normal to the surface. This setup is actually realized in the AdS/CFT, and we have already examined in Refs.~\cite{Matsueda2}. Then, the integrated entropy value $S_{\Sigma}$ is given by
\begin{eqnarray}
S_{\Sigma} = -\int_{\Sigma}\theta^{\alpha}d\eta_{\alpha} = -\int_{\Sigma}\theta^{\alpha}n_{\alpha}\frac{d\Sigma}{2\pi} = -\frac{1}{2\pi}\int_{\Psi}\nabla_{\alpha}\theta^{\alpha}d\Psi,
\end{eqnarray}
where $\Psi$ is a region surrounded by the boundary $\Sigma$ and $2\pi$ appears after integration of rotational degree of freedom in a polar coordinate for $d\Sigma$. The result means that the bulk information flow does cancel out, and only the boundary flow remains.

\subsection{Derivation of AdS${}_{d+2}$ metric and entanglement entroy for CFT${}_{d+1}$ from Hessian potential}

The Hessian potential that exactly produces AdS${}_{d+2}$ is given by
\begin{eqnarray}
\psi(\bt) = -\kappa\ln\left(\theta-\frac{1}{2}\eta_{ij}\theta^{i}\theta^{j}\right), \label{AdS}
\end{eqnarray}
where the Roman index $i$ runs from $0$ to $d$ ($\theta^{0}=t$), and $\kappa$ is related to the central charge in $d=1$. We take the Minkowski metric $\eta_{ij}$ and the Lorenztian signature is taken to be
\begin{eqnarray}\eta_{ij}\theta^{i}\theta^{j}=-(\theta^{0})^{2}+\sum_{a=1}^{d}(\theta^{a})^{2}.
\end{eqnarray}
We define the new coordinates as
\begin{eqnarray}
z=\sqrt{\theta-\frac{1}{2}\eta_{ij}\theta^{i}\theta^{j}} \; , \; 
x^{i}=\frac{1}{2}\theta^{i} \; (i=0,1,...,d).
\end{eqnarray}
This coordinate transformation leads to the Poincar\'{e}-disk form of the AdS${}_{d+2}$ spacetime. The domain of $\psi$ is given by
\begin{eqnarray}
\theta-\frac{1}{2}\eta_{ij}\theta^{i}\theta^{j}>0.
\end{eqnarray}
By using $z$ and $x^{i}$, the Fisher metric is evaluated as follows
\begin{eqnarray}
g = g_{\mu\nu}d\theta^{\mu}d\theta^{\nu} = 4\kappa\frac{dz^{2}+\eta_{ij}dx^{i}dx^{j}}{z^{2}}.
\end{eqnarray}
We may multiply a length scale of the curvature radius of AdS to $g$ to introduce an appropriate physical unit, since $\kappa$ corresponds to the Brown-Henneaux central charge in the $d=1$ case. Thus, we introduce a line element squared as
\begin{eqnarray}
ds^{2}=\frac{R^{2}}{4\kappa}g,
\end{eqnarray}
where $R$ is the curvature radius of the AdS spacetime.

Substituting Eq.~(\ref{AdS}) to Eq.~(\ref{exact}) directly leads to the entanglement entropy formula. In the case of logarithmic Hessian potential, the entropy is also logarithmic and we find
\begin{eqnarray}
S=-\kappa\ln\left(\theta-\frac{1}{2}\eta_{ij}\theta^{i}\theta^{j}\right)+\kappa\frac{\theta-\eta_{ij}\theta^{i}\theta^{j}}{\theta-\frac{1}{2}\eta_{ij}\theta^{i}\theta^{j}}.
\end{eqnarray}
Changing the notation with use of the original model parameters such as subsystem size $L$, we obtain the explicit formula of the entanglement entropy. The most important parameter for the entropy scaling is $\theta$ that controls a length scale emerging from truncation of environmental degrees of freedom.

In a spacially one-dimensional case ($d=1$), we have found in Refs.~\cite{Matsueda1,Matsueda2}
\begin{eqnarray}
\theta=L^{-2}.
\end{eqnarray}
The entanglement entropy is then given by $S\simeq 2\kappa\ln L$, and for $\kappa=c/6$ with the central charge $c$ this is consistent with the logarithmic entropy formula in CFT${}_{1+1}$~\cite{Holzhey,Calabrese1,Calabrese2,Brown}. For this examination, we have started with spinless free fermions on a discretized lattice, $\mathcal{H}=-t\sum_{i}\left(c_{i}^{\dagger}c_{i+1}+c_{i+1}^{\dagger}c_{i}\right)$, which is in a non-relativistic case. There, the band dispersion was given with the Fourier transformation by $\epsilon_{k}=-2t\cos k\sim -2t(1-k^{2}/2)$, and we gusss that the bilinear feature of the dispersion reflects on the entanglement spectrum even after taking truncation of environmental degrees of freedom. Thus, the exponent may depend on such a setup. If we consider a relativisrtic model, we expect that $\theta=L^{-1}$.

On the other hand, higher-dimensional cases seem to be characterized by
\begin{eqnarray}
\theta=\exp\left(-(a/\kappa)L^{d-1}\right), \label{theta2d}
\end{eqnarray}
and then we find the area-law scaling $S\simeq aL^{d-1}$ with a constant factor $a$. Although we do not still confirm it theoretically, this would be related to difficulty of numerical simulations of quantum critical models for $d>1$. This is because Eq.~(\ref{theta2d}) shows extremely dense entanglement energy levels. In the density matrix renormalization group method and the theory for finite-entanglement scaling~\cite{White1,White2,Pollmann,Lefevre}, the truncation number $\chi$ in the Schmidt decomposition in Eq.~(\ref{Schmidt}), $\left|\psi\right>\simeq\left|\psi_{\chi}\right>=\sum_{n=1}^{\chi}\sqrt{\lambda_{n}}\left|n\right>\otimes\left|\tilde{n}\right>$ with $\lambda_{1}\ge\lambda_{2}\ge\cdots\ge\lambda_{\chi}$, should be taken so that the appropriate entropy scaling appears. In this case, $S\sim\zeta\log\chi$ with a constant $\zeta$ owing to geometric partition with $\chi$ degrees of freedom, and then
\begin{eqnarray}
\chi\sim e^{S/\zeta}\sim \exp\left((a/\zeta)L^{d-1}\right).
\end{eqnarray}
The magnitude of $\chi$ is related to how many states should be kept in $\left|\psi_{\chi}\right>$ for reasonable approximation. This suggests that the inverse of $\chi$ roughly determines the spacing of the entanglement energy levels. Thus, it is reasonable to consider that the corresponding $\theta$ is given by $\theta\sim \chi^{-1}$ and $\kappa\sim\zeta$.

\section{Hessian potential deformed from pure AdS}

\subsection{Extention of Eq.~(\ref{AdS})}

The presence of the FG term can be represented by adding a term into Eq.~(\ref{AdS}) as follows:
\begin{eqnarray}
\Psi_{h}(\bx) = -\kappa\ln\left(\theta-\frac{1}{2}\eta_{ij}\theta^{i}\theta^{j}-h(\bt)\right),
\end{eqnarray}
where $h$ is a function of the canonical parameters $\bt$. The parallel description of $h$ with $\eta_{ij}\theta^{i}\theta^{j}/2$ indicates that the second derivative of $h$ gives a purturbation from the Minkowski metric at the AdS boundary where the CFT${}_{d+1}$ lives. We introduce new coordinates as
\begin{eqnarray}
z=\sqrt{\theta-\frac{1}{2}\eta_{ij}\theta^{i}\theta^{j}-h(\bt)} \; , \; x^{i}=\frac{1}{2}\theta^{i} \; (i=0,1,...,d).
\end{eqnarray}
The Fisher metric is given by
\begin{eqnarray}
G_{\mu\nu}[h]=\partial_{\mu}\partial_{\nu}\Psi_{h}(\bt).
\end{eqnarray}
We expand the metric to a power series of $h$
\begin{eqnarray}
G[h]=g+\kappa\left\{ \Delta K_{1}[h] + \Delta K_{2}[h] + \cdots \right\}
\end{eqnarray}
where $G[0]=g$ and we use the abbreviation $G=G_{\mu\nu}d\theta^{\mu}d\theta^{\nu}$ and $\Delta K_{i}=(\Delta K_{i})_{\mu\nu}d\theta^{\mu}d\theta^{\nu}$. We focus on the first order perturbation $\Delta K_{1}[h]$.

\subsection{Classical side}

The full representation of the Fisher metric including $h$ is given by
\begin{eqnarray}
\frac{G}{\kappa} &=& \left[ \frac{\partial\partial h}{z^{2}}+\frac{(1-\partial h)^{2}}{z^{4}}\right]\left(2zdz+4\eta_{kl}x^{k}dx^{l}+dh\right)^{2} \nonumber \\
&& + 4\left[\frac{\partial_{i}\partial h}{z^{2}}-\frac{(1-\partial h)(2\eta_{ij}x^{j}+\partial_{i}h)}{z^{4}}\right]dx^{i}\left(2zdz+4\eta_{kl}x^{k}dx^{l}+dh\right) \nonumber \\
&& + 4\left[\frac{\eta_{ij}+\partial_{i}\partial_{j}h}{z^{2}}+\frac{(2\eta_{ik}x^{k}+\partial_{i}h)(2\eta_{jl}x^{l}+\partial_{j}h)}{z^{4}}\right]dx^{i}dx^{j}.
\end{eqnarray}
We are careful for the definition that the derivative $\partial_{\alpha}$ is taken by $\theta^{\alpha}$, not $z$ and $x^{i}$. The kernel proportional to $h$ is given by
\begin{eqnarray}
\Delta K_{1}[h] &=& 4(\partial\partial h)(dz)^{2}+16\frac{(\partial\partial h)}{z}dz\eta_{ij}x^{i}dx^{j} +16\frac{(\partial\partial h)}{z}\eta_{il}\eta_{jk}x^{i}x^{j}dx^{k}dx^{l} \nonumber \\
&& +8\frac{(\partial_{i}\partial h)}{z}dzdx^{i}+16\frac{(\partial_{i}\partial h)}{z^{2}}dx^{i}\eta_{kl}x^{k}dx^{l}+4\frac{(\partial_{i}\partial_{j}h)}{z^{2}}dx^{i}dx^{j}.
\end{eqnarray}
Note again that $\partial=\partial/\partial\theta=\partial/\partial\theta^{d+1}$. If $x^{i}$ are small enough (this assumption is also used in related works), we can approximately take
\begin{eqnarray}
\Delta K_{1}[h] \simeq 4(\partial\partial h)(dz)^{2}+8\frac{(\partial_{i}\partial h)}{z}dzdx^{i}+4\frac{(\partial_{i}\partial_{j}h)}{z^{2}}dx^{i}dx^{j}.
\end{eqnarray}
As already mentioned, the term $h$ seems to be a source potential of the deformation of the boundary Fisher metric. This means that the Hessian structure holds even at the AdS boundary. According to the definition of the FG term, we see
\begin{eqnarray}
\partial_{i}\partial_{j}h=z^{d+1}H_{ij} \; , \; \partial\partial h=z^{d-1}H \; , \; \partial_{i}\partial h=0. \label{hh}
\end{eqnarray}
Thus, we actually notice that the $h$ term is really transformed into the FG term.

With respect to the holographic stress tensor in Eq.~(\ref{TH}), it is helpful to translate the energy-momentum conservation law to a relation for the boudary Hessian potential $h$. Since the Hessian structure is still kept at the AdS boundary, the conservation relation can be simply described. Let us start with
\begin{eqnarray}
\partial_{0}T^{00}+\partial_{a}T^{0a}=0,
\end{eqnarray}
where $a=1,2,...,d$. Combining this with Eq.~(\ref{hh}), we find $\partial_{0}\partial^{0}\partial^{0}h+\partial_{a}\partial^{0}\partial^{a}h=0$, and then
\begin{eqnarray}
\partial_{0}\partial^{0}h+\partial_{a}\partial^{a}h=0. \label{conserve}
\end{eqnarray}
In general a time-independent factor remains in the right hand side, but here we neglect it. At the same time, even when we start with $\partial_{0}T^{0b}+\partial_{a}T^{ab}=0$, we obtain the same result. This equation will be used for the evaluation of the holographic entanglement entropy. Note that using the energy-momentum conservation law is one of physical constraints in the present approach.

\subsection{Quantum side}

According to Eq.~(\ref{exact}), the entanglement entropy with $h$ is obtained as
\begin{eqnarray}
S_{h} = -\kappa\ln\left(\theta-\frac{1}{2}\eta_{ij}\theta^{i}\theta^{j}-h\right) + \kappa\frac{\theta-\eta_{ij}\theta^{i}\theta^{j}-\theta^{\alpha}\partial_{\alpha}h}{\theta-\frac{1}{2}\eta_{ij}\theta^{i}\theta^{j}-h}. \label{Sh}
\end{eqnarray}
We expand $S$ as a power series of $h/z^{2}$ as
\begin{eqnarray}
S_{h}=S_{0}+\Delta S_{1}[h/z^{2}] + \Delta S_{2}[h/z^{2}] + \cdots,
\end{eqnarray}
and pick up the first order $\Delta S_{1}$. We find
\begin{eqnarray}
\Delta S_{1} = \kappa\frac{s_{h}}{z^{2}},
\end{eqnarray}
where we have defined
\begin{eqnarray}
s_{h}=h-\theta^{\alpha}\partial_{\alpha}h.
\end{eqnarray}
This quantity $s_{h}$ is directly related to the entropy change at the boundary, and holds the Hessian structure. We differentiate $\Delta S_{1}$ by $\theta^{j}$ and obtain
\begin{eqnarray}
\partial_{j}\Delta S_{1} = \kappa\frac{-\theta^{i}\partial_{i}\partial_{j}h}{z^{2}} = -\kappa\theta^{i}z^{d-1}H_{ij}.
\end{eqnarray}
Here we neglect $\theta\partial\partial_{j}h$ at the AdS boundary. This substitution of Eq.~(\ref{hh}) is crucial for the derivation of the entropy-energy relation. Since the entanglement entropy formula in Eq.~(\ref{Sh}) has the deformation factor $h$, this feature roughly corresponds to the minimal surface change in the application method of the Ryu-Takayanagi formula.

If $H_{ij}$ are constants, we can perform integration as
\begin{eqnarray}
\Delta S_{1} = -\kappa\int z^{d-1}\theta^{i}H_{ij}d\theta^{j} = 2\kappa z^{d-1}\left(-x^{i}H_{ij}x^{j}\right).
\end{eqnarray}
This result can be further transformed into a simpler form by using the energy-momentum conservation relation at the boundary. By using Eq.~(\ref{conserve}), we find
\begin{eqnarray}
-x^{i}H_{ij}x^{j}=-x_{0}H^{0}_{\; 0}x^{0}-x_{a}H^{a}_{\; b}x^{b}\simeq m\left(-x_{0}x^{0}+x_{a}x^{a}\right)=m\sum_{i=0}^{d}(x^{i})^{2},
\end{eqnarray}
where $a,b=1,2,...,d$ and $m=H^{0}_{\; 0}=-H^{i}_{\; i}$. Fortunately, we have obtained the ball-shaped volume automatically. This setup is similar to the previous results. As already examined, if we consider a ball-shaped region $A$, it is enough to use only $H_{00}$. Usually we fix time as $t=0$ for the calculation of the entanglement entropy. Thus, we conclude
\begin{eqnarray}
\Delta S_{1} = 2\kappa z^{d-1}m\sum_{a=1}^{d}(x^{a})^{2}.
\end{eqnarray}
The term $m\sum_{a=1}^{d}(x^{a})^{2}$ can be identified with $\Delta E_{A}$ except for some factors. This is a well-known form of the entropy-energy relation. Therefore, our starting point, Eq.~(\ref{exact}), is a very fundamental thermodynamic relation even for the theory of quantum entanglement. In cases of $d>1$, the factor $z^{d-1}$ remains. we can evaluate it as $z^{d-1}\sim\theta^{(d-1)/2}\sim\exp\left(-(d-1)(a/2\kappa)L^{d-1}\right)$. This means that the entanglement energy level is dense enough in higher-dimensional cases and the entropy difference between the ground and first-excited states is very small.

Let us look at an example. The CFT${}_{2}$ is the simplest case. The entropy is given by
\begin{eqnarray}
\Delta S_{1} = \frac{c}{3}m(x^{1})^{2}=\frac{c}{3}m\left(\frac{l}{L}\right)^{2},
\end{eqnarray}
where $\kappa=c/6$ according to the scaling formula of the entanglement entropy and $l(\bar{n})$ is a function of filling fraction $\bar{n}$. The structure of this result is quite similar to the CFT one by Alcaraz~\cite{Alcaraz}.

\section{Summary and remarks}

In this study, we have examined the entanglement thermodynamics in viewpoints of the Fefferman-Graham metric as well as the Hessian geometry. The most important finding is that the deformation of the bulk Hessian potential from the pure AdS background behaves as a souce potential of the boundary Fisher metric, and thus the deformation coincides with the Fefferman-Graham term. By this method, we can relate the general formula of entanglement thermodynamics by the Hessian geometry to the results from quantum field theory side. We need not to start with the Ryu-Takayanagi formula. This means that the Hessian-potential approach contains enough information of the area law formula of the entanglement entropy. It is clear that the presence of $h$ changes the shape of the entangling surface and the domain boundary, and thus the similarity between the present approach and the Ryu-Takayanagi formula is quite natural.

At the same time, our parameter space is not a real spacetime, and it is still mysterious whether the present approach and the standard AdS/CFT correspondence are related with each other. It would be a future step to understand the meaning of this structural similarity of both theories.

I am interested in whether the black hole metric is exactly represented by the Hessian potential. In my previous work~\cite{Matsueda4}, it was possible to introduce the Fisher metric that leads to both of the BTZ black hole and the finite-temeprature version of the entanglement entropy formula. However, this derivation contains an approximation in which we replace $\sinh(z/z_{0})$ by $\sin^{-1}(z/z_{0})$ in a case of $z\ll z_{0}$. We may resolve this problem by the Hessian potential approach.

\acknowledgments
This work was supported by JSPS KAKENHI Grant Number 15K05222.


\begin{thebibliography}{99}

\bibitem{Matsueda1}
Hiroaki Matsueda, \emph{Geometry and Dynamics of Emergent Spacetime from Entanglement Spectrum}, arXiv:1408.5589[hep-th].
\bibitem{Matsueda2}
Hiroaki Matsueda, \emph{Hessian geometry and entanglement thermodynamics}, arXiv:1508.02538[hep-th].
\bibitem{Matsueda3}
Hiroaki Matsueda, \emph{Correspondence between causality in flat Minkowski spacetime and entanglement in thermofield-double state: Hessian geometrical study}, arXiv:1508.04679[hep-th].


\bibitem{Casini}
Horacio Casini, Marina Huerta, and Robert C. Myers, \emph{Towards a derivation of holographic entanglement entropy}, \emph{JHEP} {\bf 05} (2011) 036.
\bibitem{Guo}
Wu-zhong Guo, Song He, and Jun Tao, \emph{Note on Entanglement Temperature for Low Thermal Excited Statess in Higher Derivative Gravity}, \emph{JHEP} {\bf 08} (2013) 050.
\bibitem{Blanco}
David D. Blanco, Horacio Casini, Ling-Yan Hung, and Robert C. Myers, \emph{Relative entropy and holography}, \emph{JHEP} {\bf 08} (2013) 060.
\bibitem{Alishahiha}
Mohsen Alishahiha, Davood Allahbakhshi, and Ali Naseh, \emph{Entanglement Thermodynamics}, \emph{JHEP} {\bf 08} (2013) 102.
\bibitem{Wong}
Gabriel Wong, Israel Klich, Leopoldo A. Pando Zayas, and Diana Vaman, \emph{Entanglement temperature and entanglement entropy of excited states}, \emph{JHEP} {\bf 12} (2013) 020.
\bibitem{Takayanagi2}
Jyotirmoy Bhattacharya, Masahiro Nozaki, Tadashi Takayanagi, and Tomonori Ugajin, \emph{Thermodynamical Properties of Entanglement Entropy for Excited States}, \emph{Phys. Rev. Lett.} {\bf 110} (2013) 091602.
\bibitem{Takayanagi3}
Masahiro Nozaki, Tokiro Numasawa, Andrea Prudenziati, and Tadashi Takayanagi, \emph{Dynamics of Entanglement Entropy from Einstein Equation}, \emph{Phys. Rev. D} {\bf 88} (2013) 026012.
\bibitem{Takayanagi4}
Jyotirmoy Bhattacharya and Tadashi Takayanagi, \emph{Entropic counterpart of perturbative Einstein equation}, \emph{JHEP} {\bf 10} (2013) 219.
\bibitem{Faulkner}
Thomas Faulkner, Monica Guica, Thomas Hartman, Robert C. Myers, and Mark Van Raamsdonk, \emph{Gravitiation from entanglement in holographic CFTs}, \emph{JHEP} {\bf 03} (2014) 051.
\bibitem{Nima}
Nima Lashkari, Michael B. McDermott, Mark Van Raamsdonk, \emph{Gravitational dynamics from entanglement "thermodynamics"}, \emph{JHEP} {\bf 04} (2014) 195.
\bibitem{Banerjee}
Shamik Banerjee, Arpan Bhattacharyya, Apratim Kaviraj, Kallol Sen, and Aninda Sinha, \emph{Constraining gravity using entanglement in AdS/CFT}, \emph{JHEP} {\bf 05} (2014) 029.
\bibitem{Arpan1}
Arpan Bhattacharyya and Aninda Sinha, \emph{Entanglement entropy from the holographic stress tensor}, arXiv:1303.1884[hep-th].
\bibitem{Arpan2}
Arpan Bhattacharyya and Aninda Sinha, \emph{Entanglement entropy from surface terms in general relativity}, arXiv:1305.3448[hep-th].


\bibitem{Maldacena1}
J. M. Maldacena, \emph{The Large N Limit of Superconformal Field Theories and Supergravity}, \emph{Adv. Theor. Math. Phys.} {\bf 2} (1998) 231.
\bibitem{Maldacena2}
O. Aharony, S. S. Gubser, J. M. Maldacena, H. Ooguri, and Y. Oz, \emph{Large-N field theories, string theory and gravity}, \emph{Phys. Rep.} {\bf 323} (2000) 183.

\bibitem{Amari}
Shun-ichi Amari and Hiroshi Nagaoka, \emph{Methods of Information Geometry}, Oxford (2000).
\bibitem{Shima}
Hirohiko Shima, \emph{Hessian Goemetry}, Shokabo, Tokyo (2001).
\bibitem{Balian}
Roger Balian, \emph{The Entropy-Based Quantum Metric}, \emph{Entropy} {\bf 16} (2014) 3878.
\bibitem{Barbaresco}
Fr\'{e}d\'{e}ric Barbaresco, \emph{Koszul Information Geometry and Souriau Geometric Temperature/Capacity of Lie Group Thermodynamics}, \emph{Entropy} {\bf 16} (2014) 4521.

\bibitem{Takayanagi1}
Shinsei Ryu and Tadashi Takayanagi, \emph{Holographic Derivation of Entanglement Entropy from the anti-de Sitter Space/Conformal Field Theory}, \emph{Phys. Rev. Lett.} {\bf 96} (2006) 181602.


\bibitem{Balasubramanian}
V. Balasubramanian and P. Kraus, \emph{A Stress Tensor For Anti-de Sitter Gravity}, \emph{Comm. Math. Phys.} {\bf 208} (1999) 413.
\bibitem{Haro}
S. de Haro, S. N. Solodukhin, and K. Skenderis, \emph{Holographic Reconstruction of Spacetime and Renormalization in the AdS/CFT Correspondence}, \emph{Comm. Math. Phys.} {\bf 217} (2001) 595.

\bibitem{Blau}
M. Blau, K. S. Narain, and G. Thompson, \emph{Instantons, the Information Metric, and the AdS/CFT Correspondence}, arXiv:0108122[hep-th].
\bibitem{Shock}
Emanuel Malek, Jeff Murugan, and Jonathan P. Shock, \emph{The Information Metric on the moduli space of instantons with global symmetries}, arXiv:1507.08894[hep-th].

\bibitem{Holzhey}
Christoph Holzhey, Finn Larsen, Frank Wilczek, \emph{Geometric and Renormalized Entropy in Conformal Field Theory}, \emph{Nucl. Phys. B} {\bf 424} (1994) 443.
\bibitem{Calabrese1}
P. Calabrese and J. Cardy, \emph{Entanglement entropy and quantum field theory}, \emph{J. Stat. Mech.} {\bf 0406} (2004) P06002 [note added: arXiv:hep-th/0405152].
\bibitem{Calabrese2}
Pasquale Calabrese and John Cardy, \emph{Entanglement entropy and conformal field theory}, \emph{J. Phys. A} {\bf 42} (2009) 504005.
\bibitem{Brown}
J. D. Brown and M. Henneaux, \emph{Cental charges in the canonical realization of asymptotic symmetries: an example from three-dimensional gravity}, \emph{Comm. Math. Phys.} {\bf 104} (1986) 207.

\bibitem{White1}
Steven R. White, \emph{Density matrix formulation for quantum renormalization groups}, \emph{Phys. Rev. Lett.} {\bf 69} (1992) 2863.
\bibitem{White2}
Steven R. White, \emph{Density-matrix algorithms for quantum renormalization groups}, \emph{Phys. Rev. B} {\bf 48} (1993) 10345.
\bibitem{Pollmann}
Frank Pollmann, Subroto Mukerjee, Ari M. Turner, and Joel Moore, \emph{Theory of Finite-Entanglement Scaling at One-Dimensional Quantum Critical Points}, \emph{Phys. Rev. Lett.} {\bf 102} (2009) 255701.
\bibitem{Lefevre}
P. Calabrese and Lefevre, \emph{Entanglement spectrum in one-dimensional systems}, \emph{Phys. Rev. A} {\bf 78} (2008) 032329.

\bibitem{Alcaraz}
Francisco Castilho Alcaraz, Miguel Ib\'{a}$\tilde{\rm n}$ez Berganza, and Germ\'{a}n Sierra, \emph{Entanglement of Low-Energy Excitations in Conformal Field Theory}, \emph{Phys. Rev. Lett.} {\bf 106} (2011) 201601.

\bibitem{Matsueda4}
Hiroaki Matsueda, \emph{BTZ Black Hole in Fisher Information Spacetime}, arXiv:1409.3908[hep-th].

\end{thebibliography}
\end{document}